\documentclass[twoside]{article}

\makeatletter
\renewcommand{\@oddhead}%
 {\raisebox{0pt}[\headheight][0pt]%
  {\vbox%
   {\hbox to \textwidth{\hfil%
    {\small\it Quantum geometrodynamics in extended phase
               space -- I}%
    \strut \hfil}%
   \hrule}}}
\renewcommand{\@evenhead}%
 {\raisebox{0pt}[\headheight][0pt]%
  {\vbox%
   {\hbox to \textwidth{\hfil%
    {\small\it V.A. Savchenko, T.P. Shestakova and G.M. Vereshkov}%
    \strut \hfil}%
   \hrule}}}
\renewcommand{\@oddfoot}{\hfil \thepage \hfil}
\renewcommand{\@evenfoot}{\hfil \thepage \hfil}

\newcounter{sect}
\newcommand{\sect}[1]%
 {\par\vspace*{4mm plus 1mm minus .5mm}\refstepcounter{sect}
  \begin{center}
   \large\bf \thesect.\hspace{2pt}#1
   \samepage
  \end{center}}
\renewcommand{\thesect}{\arabic{sect}}

\makeatother

\begin{document}

\thispagestyle{plain}
\renewcommand{\thefootnote}{\fnsymbol{footnote}}

\begin{center}
{\Large\bf QUANTUM GEOMETRODYNAMICS\\[2pt]
           IN EXTENDED PHASE SPACE -- I.\\[10pt]
{\large\sl PHYSICAL PROBLEMS OF INTERPRETATION\\[2pt]
           AND MATHEMATICAL PROBLEMS OF GAUGE INVARIANCE}

\vspace{10mm}
V. A. Savchenko\footnote{e-mail: savchenko@phys.rnd.runnet.ru}
T. P. Shestakova\footnote{e-mail: stp@phys.rnd.runnet.ru}
and G. M. Vereshkov}

\vspace{5mm}
{\it Dept. of Theoretical Physics, Rostov State University,
     Sorge Str. 5, Rostov-on-Don, Russia}
\end{center}

\begin{abstract}
\small
The paper is the first of two parts of the work devoted to the investigation
of constructing quantum theory of a closed universe as a system without
asynptotic states. In Part I the role of asymptotic states in quantum theory
of gravity is discussed, that enables us to argue that mathematically
correct quantum geometrodynamics of a closed universe has to be
gauge-noninvariant. It is shown that a gauge-noninvariant quantum
geometrodynamics is consistent with the Copenhagen interpretation.
The proposed version of the theory is thought of as describing the
integrated system ``the physical object + observation means''. It is also
demonstrated that introducing the observer into the theory causes the
appearance of time in it.
\end{abstract}

\begin{center}
\end{center}
\date{}

\setcounter{footnote}{0}
\renewcommand{\thefootnote}{\arabic{footnote}}

\sect{Introduction}
\label{Introduction}
The purpose of this work is to explore a possibility of constructing
physically (operationally) interpreted quantum geometrodynamics (QGD) of
a closed universe by a strict mathematical method without using any
assumption not permitting detailed mathematical proofs.

The Wheeler -- DeWitt QGD \cite{DeWitt} is the extrapolation of conceptions
and methods of modern quantum field theory. The validity of them on the
scale of the Universe as a whole may arouse doubts. In our paper we attract
a special attention to the fact that a closed universe is a system without
asymptotic states, in contrast to those normally considered by quantum field
theory. In our opinion, taking into account this fact may lead to a theory
significantly distinguish from the Wheeler -- DeWitt QGD. Namely, we have
come to the conclusion that mathematically correct and physically
well-grounded QGD of a closed universe is a gauge-noninvariant theory.

The grounds for this conclusion are discussed in the presented below first
of the two papers. We show that any gauge-invariant quantum field
theory is essentially based on the assumption about asymptotic states.
Indeed, to construct a gauge-invariant theory one needs 1) to pass to
a description of the system under consideration in terms of ``true''
physical degrees of freedom, or 2) to impose some selection rules singling
out physical state vectors. The first way implies that constraints have to
be resolved. For gravitational constraints the latter can be done in the
limits of perturbation theory only in asymptotically flat spaces, or in
some special cases. As for the second way, in the path integral
approach which is adopted in the present investigation as more adequate,
selection rules for physical states are equivalent to asymptotic boundary
conditions in a path integral, so if the path integral is considered without
asymptotic boundary conditions (as it should be in a correct quantum theory
of a closed universe) the set of all possible transition amplitudes
determined through the path integral inevitably involves gauge-noninvariant
ones.

On the basis of the Copenhagen operational interpretation of quantum theory
(QT), we establish the discrepancy of the mathematical structure of the
gauge-invariant theory to the conditions of observations in a closed
universe. In contrast, from the viewpoint of the Copenhagen interpretation,
it should be expected that a wave function of the Universe must carry
information on geometry of the Universe as well as on a reference system
(RS) in which this geometry in studied and which represents the observer
in the theory of gravity. All this give rise to the idea of a
gauge-noninvariant quantum geometrodynamics of a closed universe.
The present work should be thought of as an attempt to give a
self-consistent description of a physical object (the Universe) and
observation means (a reference system).

The wave function that carries information on geometry of the Universe and
on the observer is supposed to satisfy a Schr\"odinger equation. The path
integral approach contains the procedure of derivation of the Schr\"odinger
equation from the path integral in a Lagrangian form and does not require
to construct a Hamiltonian form of the theory. For gravity appropriate
calculations can be made in the class of differential gauge conditions
introducing missing velocities into the Lagrangian. All the calculations are
demonstrated explicitly in the second part of our paper taking the Bianchi
IX cosmological model as an example.

To approximate the path integral when deriving the Schr\"odinger equation
we use the ``extended'' set of Lagrangian equations including those for
ghosts and a gauge condition. The set of equations can be obtained by
varying the Batalin -- Vilkovisky (BV) effective action (which reduces to
the Faddeev -- Popov effective action in the present case). Let us note
that at this point the Batalin -- Vilkovisky (Lagrangian) formalism
\cite{BV} turns out not to be equivalent to the Batalin -- Fradkin --
Vilkovisky (Hamiltonian) one \cite{BFV1, BFV2, BFV3, BFV4} in the sense
that the set of Hamiltonian equations in extended phase space (EPS)
obtained from the Batalin -- Fradkin -- Vilkovisky (BFV) action is not
equivalent to the extended Lagrangian set of equations. The reason is a
different structure of ghost sectors, which, in its turn, results from
the fact that the gauge group of gravity does not coincide with the group
of canonical transformations generated by gravitational constraints. In the
case of a system with asymptotic states the ghost sector plays just an
auxiliary role and it is possible to prove the equivalence of the
Lagrangian and Hamiltonian schemes in a gauge-invariant sector. However,
in the case of a closed universe we are not able, in principle, to single
out a gauge-invariant sector by means of asymptotic boundary conditions,
and it is clear from general consideration that making use of non-equivalent
sets of equations leads to different equations for the wave function of the
Universe.

In this situation we give preference to the Lagrangian formulation of the
theory. At the same time in the class of gauges mentioned above a
Hamiltonian formulation equivalent to the Lagrangian one can be constructed.
BRST-transformations in EPS then correspond to gauge transformations of
the original theory rather than those generated by the constraints.

In our formulation of Hamiltonian dynamics in EPS the passage to a
gauge-invariant theory can be made in a usual manner: all gauge-noninvariant
terms in the set of equations in EPS can be excluded by means of asymptotic
boundary conditions%
\footnote{Here we ignore the related problem of Gribov's copies, which,
in our opinion, needs a special consideration in this context.}.
One then will come to gauge-invariant Hamiltonian equations complemented
by Dirac's primary and secondary constraints. Thus, from the viewpoint
which we advocate here, the Dirac formulation of constrained dynamics seems
to correspond to a particular situation when a gauge-invariant sector can
be singled out, and so does the BFV approach which inherits many features
of Dirac's scheme. The requirement of BRST-invariance of a state vector
leads to the Wheeler -- DeWitt equation in the BFV approach but not in
the approach to constructing Hamiltonian dynamics in EPS presented here.

In our approach the wave function satisfying the Wheeler -- DeWitt equation
is a particular solution to the Schr\"odinger equation answering a special
choice of a gauge. Furthermore, bearing in mind that the choice of
parametrization of gauge variables and the choice of gauge conditions
together fix a reference system, we come to the conclusion that the
well-known parametrization noninvariance of the Wheeler -- DeWitt equation
is in essence an ill-hidden gauge noninvariance. So the Wheeler -- DeWitt
QGD cannot be thought of as a gauge-invariant theory in a strict sense.

Our papers are organized as following.

In Sec.\,\ref{Many-worlds} we briefly remember the paradigm on which the
Wheeler -- DeWitt QGD is based, and in Sec.\,\ref{Algorithm} we formulate
the algorithm of our research. In Sec.\,\ref{Asymp.states} the role of
asymptotic states in quantum theory of gravity is discussed that leads us
to the conclusion that quantum theory of a closed universe, in general, may
be not gauge-invariant. In Sec.\,\ref{Integrity} we show that a
gauge-noninvariant QGD is compatible with the Copenhagen interpretation of
QT. The interpretation of a reference system fixing the whole information
on the evolution of the Universe is given in Sec.\,\ref{RS_interpret}. The
choice of an appropriate formalism for constructing mathematically correct
QGD of a closed universe is motivated in Sec.\,\ref{scheme}. In
Sec.\,\ref{Lagr.eqs} concluding the first part of the work the
gauge-noninvariant extended set of Lagrangian equations is considered,
and it is shown that time-ordering in quantum dynamics is a consequence
of introducing the observer into QGD equations.

The detailed mathematical proofs of the possibility to realize the proposed
approach are given in the second part of the work. All the mathematical
operations concerned with deducing a gauge-noninvariant Schr\"odinger
equation and constructing its general solution are carried out in a
manifest form for the Bianchi-IX model.

The Bianchi IX cosmological model has been chosen for its mathematical
simplicity and physical meaningfulness. It is traditionally used
as a test polygon for various theoretical methods in cosmology (see, for
example, \cite{Duncan}). The finite number of degrees of freedom enables
us to control correctness of  used methods under the ``pure'' conditions
without mathematical problems connected to divergences typical for a
general quantum field theory with infinite number of degrees of freedom.

In our opinion the approach to QGD of a closed universe proposed by us
phenomenologically demonstrates the existence of the problem of searching
new fundamental physical principles describing the process of forming the
integrated system ``a physical object + observation means'' and controlling
measuring processes in this system and mechanisms of registration of
measurement results on physical carriers of measuring devices. Some aspects
of this problem are discussed in Conclusions.

\sect{The many-worlds interpretation of quantum geometrodynamics}
\label{Many-worlds}
The standard QGD is based on the Wheeler -- DeWitt equations \cite{DeWitt}
\begin{equation}
\label{WDWeqns}
{\cal T}^{\mu}|\Psi\rangle =0,
\end{equation}
where the operators
\begin{equation}
\label{T_constr}
{\cal T}^0 =
 (-\gamma_{(3)})^{-\frac12}p^{ik}(\gamma_{il}\gamma_{km}
 -\frac12\gamma_{ik}\gamma_{lm})p^{lm}
 + (-\gamma_{(3)})^{\frac12}R_{(3)}+T^{00}_{(mat)},
\end{equation}
$$
{\cal T}^i=-2(\partial_kp^{ik}+\gamma^i_{lm}p^{lm})+T^{0i}_{(mat)},
$$
$p^{ik}$ are the momenta conjugate to the 3-metric $\gamma_{ik}$,
$\gamma^i_{kl}$ are the three-connections,
$\gamma_{(3)}\equiv \det\|\gamma_{ik}\|$,
$R_{(3)}$ is the 3-curvature, $T^{\mu\nu}_{(mat)}$ is the energy-momentum
tensor of the material fields. Derivation of these equations by
quantum-theoretical methods has been discussed by many authors. Because of a
number of reasons considered below and analyzed in details for the
Bianchi-IX model in Part II of our work, the Wheeler -- DeWitt
equations are not deducible by correct mathematical methods in the framework
of the ordinary quantum theory. In principle, this fact itself is not
sufficient to discard the Wheeler -- DeWitt theory. The ordinary
quantum theory is a phenomenological theory for describing quasilocal
(in a macroscopic sense) phenomena. Therefore its extrapolation to the
scales of the Universe as a whole is a radical physical hypothesis that
may be incompatible correctly with the existing formalism. In
this situation it makes sense to analyze the Wheeler -- DeWitt theory as it is, without
fixing attention on whether a correct way of its construction exists or not.

The most distinctive feature of the Wheeler -- DeWitt theory is that there is no quantum
evolution of state vector in time. Once adopting the Wheeler -- DeWitt theory, one should
admit that a wave function satisfying Eqs.\,(\ref{WDWeqns}) describes
the past of the Universe as well as its future with all observers being
inside the Universe in different stages of its evolution, and all
observations to be made by these observers. We should emphasize that the
question about the status of an observer in the Wheeler -- DeWitt theory is rather
specific since there is no vestige of an observer in Eqs.\,(\ref{WDWeqns}).
The introduction of the observer into the theory is performed by fixing
boundary conditions for a wave function of the Universe, we shall return
to them below.

First of all, one should pay attention to another one feature of the Wheeler -- DeWitt
theory: because of the status of the wave function of the Universe mentioned
above this theory does not use the postulate about {\em the reduction of
a wave packet}. The logical coordination of concepts in the Wheeler -- DeWitt theory is
carried out within the framework of the many-worlds interpretation of the
wave function proposed by Everett \cite{Everett} and applied to QGD by
Wheeler \cite{Wheeler}. The wave function satisfying Eqs.\,(\ref{WDWeqns})
and certain boundary conditions is thought to be a branch of a many-worlds
wave function that corresponds to a certain universe; other branches being
selected by other boundary conditions. Thus, the boundary conditions for the
wave function of the Universe acquire a fundamental meaning for
the theory: they fix all actions of an observer through the whole history of
the Universe, i.e. they contain the concentrated information about the
continuous reduction of the wave function in the process of evolution of
the Universe including certain observers inside \cite{Everett,Wheeler}.

Let us discuss the  peculiarities of statement of the problems in the Wheeler -- DeWitt QGD
using the Bianchi-IX model as an example, space homogeneity of the latter
reducing the set of the Wheeler -- DeWitt equations to the only equation
\begin{equation}
\label{B-IX_WDW}
H_{ph}|\Psi\rangle = 0.
\end{equation}
Gauge invariance is expressed by that the choice of a time coordinate is not
made when deriving (more precisely, when writing down) the Wheeler -- DeWitt equation. The
equation should be solved under some boundary conditions. However, carrying
out this program one should bear in mind that solutions to this equation are
unnormalizable. The latter is obvious from the following mathematical
observations: the Wheeler -- DeWitt equation coincides formally with an equation for the
eigenfunction of the physical Hamiltonian $H_{ph}$, appropriate to the zero
eigenvalue. Meanwhile, nothing prevents us from studying {\em the whole}
spectrum of eigenvalues of the operator $H_{ph}$; then the wave function
satisfying Eq.\,(\ref{B-IX_WDW}) turns out to be normalizable only if the
value $E=0$ belongs to a discrete spectrum of the operator $H_{ph}$.

As for the operator $H_{ph}$, explicit form of which will be presented in
Part II of the paper, it has a {\em continuous} spectrum. In this situation
one faces the alternative: 1) to declare the Bianchi-IX model to be
meaningless and to put the question about searching for such models, whose
operator $H_{ph}$ has a discrete spectrum line at $E = 0$, or 2) to refuse
to normalize the wave function of the Universe enlarging more, by that, the
discrepancy between QGD and ordinary quantum theory. In the Wheeler -- DeWitt QGD the
second way is chosen that does not contradict in principle to the statement about
the status of the wave function of the Universe mentioned above%
\footnote{Strictly speaking, unnormalizability of the wave function of the
Universe means that there is no probabilistic interpretation of it,
which is quite natural, because the results of  the continuous reduction
of a wave packet are involved {\em in the boundary conditions} but not in
the {\em structure of a superposition}. Nevertheless, attempts were made by
many authors to retrieve a probabilistic interpretation of QGD (see, for
example, \cite{BP}) analyzing its physical and mathematical contents by
methods of the modern gauge field theory. In such an approach mathematical
correctness of the analysis, thorough study of the question, whether the
used mathematical procedures do exist, take on fundamental significance.
Exactly that very problem will be discussed in details in our papers. The
conclusions of our investigation are: mathematically correct procedures
exist at least for systems with finite number of degrees of freedom,
however, such procedures lead to a theory radically distinguished from the
Wheeler -- DeWitt QGD by its mathematical form as well as by its physical content}.

What should the Wheeler -- DeWitt theory be taken for by an individual local observer?
Obviously -- for {\em a paradigm} fixing a certain way of thinking that
{\em in principle} cannot be verified or overthrown experimentally. The
reference to the fact that in the classical limit of the Wheeler -- DeWitt theory one can
obtain the Einstein equations, conclusions from which can be compared with
cosmological observations, is not an argument, since it is obvious in advance
that there exist an {\em infinite number of ways} to make a quantum
generalization of the classical theory of gravity based not on mathematically
correct procedures but on adopting some paradigm.

The Wheeler -- DeWitt paradigm may contain a deep sense that is inaccessible for understanding
yet. However, it is clear that its existence does not deprive another approach
to QGD problems of a sense; for instance, an approach based on adopting
another interpretative paradigm or an approach based on procedures claiming
in a greater measure for mathematical strictness then those used in the Wheeler -- DeWitt
theory.

\newpage

\sect{Our research algorithm}
\label{Algorithm}
In our work the following research algorithm is realized.
\begin{enumerate}
\item A transition amplitude between any two states of the Universe
expressed through a path integral is adopted as a basic object of QGD
(that predetermines the probabilistic interpretation of the theory).
\item We take notice of the circumstance that a closed universe has no
asymptotic states. Imposing asymptotic boundary conditions in the path
integral is not correct in this case. In this situation we do not see any
foundations to for a statement of gauge invariance of the theory.
\item Instead we state the problem of constructing a wave function of the
Universe containing information about a physical object as well as about a
RS (fixed by a gauge) in which the object is studied. The explicit solution
of the problem has been obtained for the Bianchi-IX cosmological model: this
model enables one to investigate the structure of the general solution to
the gauge-noninvariant Schr\"odinger equation.
\item We pay attention to the fact that such a wave function of the Universe
corresponding to the observation conditions in a closed universe answers the
Copenhagen interpretation of QT.
\item We take into account the notion about a RS, formulated by Landau and
Lifshitz \cite{Lan}.
\item A parameter used for fixing information about observation means (a
reference system) in the wave function of the Universe is proved to have a
mathematical status of an eigenvalue of the gravitational Hamiltonian.
\item The question of existence of the Wheeler -- DeWitt equation is
analyzed from the two positions:
a)~we find a particular solution to the dynamical Schr\"odinger equation
that corresponds to the zero eigenvalue of the gravitational Hamiltonian and
to a factored EPS; the way of deriving the Wheeler -- DeWitt equation in
this case shows, however, that the parametrization noninvariance of the
Wheeler -- DeWitt equation \cite{HP} is an ill-hidden gauge noninvariance
since any change of parametrization is equivalent to a new gauge;
b)~we consider the derivation of the Wheeler -- DeWitt equation as a
consequence of the requirement for a state vector to be BRST-invariant.
In the BFV approach the requirement of BRST-invariance leads to the
Wheeler -- DeWitt equation, but we argue that the assumption about
asymptotic states is implied in the BFV scheme.
\item For the simplified Bianchi-IX model with one of the gravitation
degrees of freedom being frozen out the exact solutions of
gauge-noninvariant conditionally-classical equations as well as the
Schr\"odinger equation are found. These solutions demonstrate explicitly
the status of an eigenvalue of the gravitational Hamiltonian as a governing
parameter that regulates properties of a cosmological solution.
\item Within the framework of the worked out QGD version we propose the
hypothesis about the Universe creation from ``Nothing'' as a reduction of a
singular state wave function.
\end{enumerate}

\sect{The assumption about asymptotic states in quantum theory of gravity}
\label{Asymp.states}
The problem of constructing quantum theory of gravity can be formulated
after adopting some postulates fixing its physical contents. As a rule such
postulates are 1) the identification of gravitational field with a system
possessing two field degrees of freedom; 2) gauge invariance of observable
quantities. In this Section we shall argue that to construct a theory
based on these two postulates one should appeal to an additional
assumption, namely, the assumption about asymptotic states. So, there is no
ground to make doubt about the legitimacy of the adopted physical postulates
in the graviton S-matrix theory. However, in a general case without
asymptotic states, which includes quantum geometrodynamics of a closed
universe, a more general approach has to be considered that may lead to
gauge-noninvariant quantum theory of gravity.

In the gauge-invariant quantum theory of gravitational field with two
degrees of freedom the correspondence principle in the most rigorous form
is adopted: the classical pre-modes of quantum equations of motion and
selection rules of physical state vectors are equations of motion and
constraints in the canonical dynamics of systems with constraints. For
a classical gravitational field the various representations of this
dynamics, which are equivalent up to canonical transformations, are offered
by Dirac \cite{Di}, Arnowitt, Deser, Misner \cite{ADM}, Faddeev \cite{Fad}.
However, it is impossible to construct an appropriate quantum theory using
canonical quantization formalism because of the degeneracy of operator
equations, since in the theory of gravity there are no canonical or any
other local gauges completely removing the degeneracy of the Einstein
equations under the diffeomorphism group transformations. The Feynman
formalism of path integration is more adequate: it allows one to control
the procedure of selection of gauge orbit representatives and contains
a method of residual degeneracy compensation.

The full set of the constraint equations ${\cal T}_{\mu}=0$ and local
gauge conditions $f^{\mu}=0$ are explicitly solvable only within the limits
of perturbation theory in asymptotically flat spaces, where the effect of
asymptotic dynamical splitting off the three-dimensionally transversal
gravitational waves from the so-called ``nonphysical" degrees of freedom
takes place. In this only case the constraint equations enable one to
express explicitly gravitational variables in terms of the ``true"
gravitational degrees of freedom. The procedure of solving the constraints
can be reproduced exactly in the framework of the path integral approach
(Faddeev and Popov, \cite{Fad, FP1, FP2}). It results in a path integral
that can be skeletonized on the extremals of the action of a system with
two field degrees of freedom. In a general case (without dynamic separation
of three-dimensionally transversal modes), which QGD of a closed universe
belongs to, the similar operations are mathematically impracticable.

In a general case one should use the ghost field technique to compensate
residual degrees of freedom not fixed by a local gauge condition. The
more powerful approach was suggested by Batalin, Fradkin and Vilkovisky
(BFV) \cite{BFV1, BFV2, BFV3, BFV4}, see also \cite{Hennaux}. The structure
of an effective action is uniquely determined by algebra of gauge
transformation, the latter, in its turn, depending on a chosen
parametrization.

In the canonical formalism the algebra of the transformations generated
by the constraints (\ref{T_constr}) is open, since one of the structure
functions in the relations
$$
\{{\cal T}_{\mu}, {\cal T}_{\nu'}\}
 =C^{\lambda''}_{\mu\nu'}{\cal T}_{\lambda''}
$$
does depend on 3-metric $\gamma_{ik}$. The BFV effective action is
\begin{equation}
\label{BFV_action}
S_{BFV}
 =\int\!\left(p^{ik}\dot\gamma_{ik}+\pi^{\mu}\dot N_{\mu}
  +{\cal P}_{\alpha}\dot\eta^{\alpha}-\{\psi,\Omega\}\right)d^4x,
\end{equation}
where $\pi^{\mu}$ are the momenta conjugate to lapse and shift functions
$N_{\mu}$; $\eta^{\alpha}, {\cal P}_{\alpha}$ are the BFV ghosts and their
momenta, in contrast to the Faddeev -- Popov ghosts
$\theta^{\mu}, \bar\theta_{\mu}$ and their momenta
$\bar\rho_{\mu}, \rho^{\mu}$;
$\psi$ is some gauge fixing function and the BRST generator is given by
\begin{equation}
\label{BRST_gen}
\Omega=\eta^{\alpha}{\cal G}_{\alpha}
  +\eta^{\alpha}\eta^{\beta}U_{\beta\alpha}^{\gamma}{\cal P}_{\gamma}
 ={\cal T}_{\mu}\theta^{\mu}-i\rho^{\mu}\pi_{\mu}
  -\frac12\bar\rho_{\mu}C^{\mu}_{\nu\lambda}\theta^{\lambda}\theta^{\nu};
\end{equation}
${\cal G}_{\alpha}$ are the full set of constraints,
$$
{\cal G}_{\alpha}=(\pi_{\mu}, {\cal T}_{\mu}),\quad
\{{\cal G}_{\alpha}, {\cal G}_{\beta'}\}
 =U^{\gamma''}_{\alpha\beta'}{\cal G}_{\gamma''}.
$$
Due to the dependence of the structure functions $C_{\mu\nu'}^{\lambda''}$
on $\gamma_{ik}$ the action (\ref{BFV_action}) will contain, in general,
a four-ghost interaction term, namely,
${\cal P}_{\alpha}\eta^{\beta}\{\chi^{\alpha},U_{\beta\delta}^{\gamma}\}
 {\cal P}_{\gamma}\eta^{\delta}$
if $\psi$ is chosen in the form $\psi={\cal P}_{\alpha}\chi^{\alpha}$
(see \cite{BFV1, BFV2, BFV4}).

At the same time, from the viewpoint of a Lagrangian approach,
the algebra of transformations of 4-metric components
\begin{equation}
\label{g_transf}
\delta g^{\mu\nu}=-\partial_{\lambda}g^{\mu\nu}\xi^{\lambda}
 +g^{\mu\lambda}\partial_{\lambda}\xi^{\nu}
 +g^{\nu\lambda}\partial_{\lambda}\xi^{\mu}
\end{equation}
is known to be closed, so that the effective action constructed in
accordance with the Batalin -- Vilkovisky method \cite{BV} will be
reduced to the Faddeev -- Popov form,
\begin{equation}
\label{FP_action}
S_{FP}
 =\int\!\left(-\frac1{2\kappa}\sqrt{-g}g^{\mu\nu}R_{\mu\nu}
  +\lambda_{\mu}f^{\mu}
  +\bar{\theta}_{\nu}\hat{M}_{\mu}^{\nu}\theta^{\mu}\right)d^4x,
\end{equation}
where $\hat{M}_{\mu}^{\nu}$ is the Faddeev -- Popov operator of the
equation for residual transformation parameters
\begin{equation}
\label{M_eta}
\hat{M}^{\nu}_{\mu}\xi^{\mu}=0;
\end{equation}
\begin{equation}
\label{FP_oper}
\hat{M}^{\nu}_{\mu}=\frac{\delta f^{\nu}}{\delta h^{\rho\sigma}}
 \left(-\partial_{\mu} h^{\rho\sigma}
  +\delta^{\rho}_{\mu}h^{\sigma\lambda}\partial_{\lambda}
  +\delta^{\sigma}_{\mu}h^{\rho\lambda}\partial_{\lambda}\right).
\end{equation}

We assume that a path integral is determined as a mathematical object
if the way of its evaluation, i. e. its skeletonization, is given.
To skeletonize the path integral one should use a full set of
extremal equations obtained by varying an appropriate effective action
including equations for ghosts and gauge conditions. Two sets of equations
obtained by varying the actions (\ref{BFV_action}) and (\ref{FP_action})
are not equivalent, and, in general, extremal equations corresponding
to various parametrizations will not be equivalent since ghost sectors
will be different. One can talk only about the equivalence of physical
(gauge-invariant) sectors. To separate a physical sector one should
consider particular solutions to extremal equations (trivial solutions
for ghosts and Lagrange multipliers of a gauge-fixing term). These
solutions, as a rule, are singled out by asymptotic boundary conditions.
The same boundary conditions ensure BRST-invariance of the path integral
and thus play the role of selection rules for physical states \cite{Hennaux}.

The full system of extremal equations is gauge-noninvariant; therefore,
the set of all transition amplitudes determined through the path integral
with the appropriate effective action will necessarily involve
gauge-noninvariant amplitudes. To select gauge-invariant amplitudes one
needs
\begin{enumerate}
\item to use the mentioned above particular solutions to the extremal
equations corresponding to trivial solutions for ghosts and Lagrange
multipliers;
\item to eliminate from the solutions of extremal equations
gauge-dependent coordinate effects existing in case of any local gauge
condition and described by the functions (\ref{g_transf}).
\end{enumerate}

The problem is that the coordinate effects can be eliminated just locally
(in the vicinity of space-time point), therefore the second operation
mentioned above can be really performed only within the limits of
perturbation theory in asymptotically flat spaces. In the other version of
perturbation theory gauge-noninvariant extremal equations are used with
the particular choice of asymptotic states (the vacuum of ghosts, 3-scalar
and 3-vector gravitons). In this case a gauge-invariant physical vector is
factored out from the state vector of the system, thus ensuring gauge
invariance of S-matrix. Hence, in asymptotically flat spaces two
gauge-invariant versions of the theory based on the path integral with
BRST-invariant effective action and the path integral evaluated on the
extremals of the action of a system with two field degrees of freedom
are equivalent by computational capabilities as well as by physical results
obtained within the limits of perturbation theory. These two versions are
therefore completely consistent.

In a general case (when transitions between nonasymptotic states are under
consideration) the procedures mentioned above cannot be carried out.
Imposing the asymptotic boundary conditions is not correct in this case.
When there are no asymptotic states we have no foundation for a statement
about gauge invariance of the theory. Strictly speaking, we even cannot
declare the path integral to be BRST-invariant since BRST-invariance
of boundary conditions is also required \cite{Hennaux} so that the boundary
conditions cannot be chosen arbitrarily. The properties of the theory should
be analyzed within this theory itself. Using gauge-noninvariant extremals
makes it clear that the answer will be gauge-noninvariant and the problem
is, whether it is possible {\em in general} to extract a gauge-invariant
information from it. Our investigation (see Part II) shows that one cannot
do it -- the fact, which we shall try to understand from the point of view
of the Copenhagen interpretation of quantum theory and, in particular, of
the Copenhagen interpretation of QGD.

\sect{Gauge noninvariance and the integrity principle}
\label{Integrity}
The Copenhagen integrity principle in its most states that a quantum
object has no properties itself; it gets and exhibits some properties
compatible with the complementarity principle only in a certain
semiclassical macroscopic situation \cite{Bohr1,Bohr2}. The main
representative of the latter is a measuring device; so, in equivalent terms,
we mean the integrity of the physical object and observation means (OM).

In quantum theory of gravity the integrity principle turns our attention to
the fact, that a gauge condition fixes a RS, i. e. it is directly related to
OM and can be operationally interpreted. Another indisputable but not less
important fact is that there exist no inertial RS in the theory of gravity.
(The special case of an island system, that can be investigated in an
asymptotically inertial RS, should be considered separately, see below). In a
general case one deals with noninertial reference systems, their properties
being represented by inertial fields. From the point of view of an observer
inside a gravitating system the existence of the inertial fields affect the
results of measurements. It is clear, that physical processes in measuring
instruments creating macroscopic background for a quantum object depend on the
inertial fields. For this reason the integrity principle does not contradict
ideologically to the assumption about specific quantum correlations between
the properties of inertial and true gravitational fields. The information
about these very correlations turns out to be contained in the
gauge-noninvariant amplitudes.

Of course, appealing to the integrity principle is a way to make a
phenomenological prediction and the question remains what is the nature of
the discussed correlations. It is obvious in advance that nonlinearity of
the gravity theory itself does not ensure the existence of unremovable
quantum gauge-noninvariant effects. An absolutely new element of physical
reality that has no analog in the classical theory must take part in their
formation. One can guess that only a nonperturbative gravitational vacuum
with broken symmetry under transformations of diffeomorphism group could be
such an element. The vacuum condensate fixing symmetry breaking must arise
as a consequence of continuous distribution of OM inside the gravitating
system. It will be shown below that this notion of gravitational vacuum can
be expressed in a mathematical language.

The operational interpretation of quantum gauge-noninvariant effects allows
to single out a particular physical situation, in which these effects should
not exist. Let us consider an island universe -- a quasilocalized coagulate
of gravitational fields -- and an observer who is about at spatial infinity
from it. The properties of the observer's RS asymptotically approaches to
the properties of an inertial system. The observer's subject of investigation
is a graviton scattering. The properties of the system ``gravitons +
detectors" ensure that the measuring devices do not affect dynamical quantum
phenomena. The role of the devices is restricted to a wave packet reduction
taking place in a space-time region asymptotically far from the
interaction region of quantum subsystems. The detectors located on bodies
forming an asymptotically inertial RS cannot fix anything but
three-dimensionally transversal gravitational waves. It is obvious that in
this situation experimental data should be described by a gauge-invariant
S-matrix. Mathematically this circumstance is taken into account by appealing
to selection rules singling out from the full set of the amplitudes
only those corresponding to observable gauge-invariant
phenomena characterized by the vacuum of ghosts, 3-scalar and 3-vector
gravitons.

The case of QGD of a closed universe is absolutely different. Firstly, in a
closed universe there is no asymptotic state in which three-dimensional
transversal gravitational waves would be dynamically split off a
longitudinal gravitational field describing the expansion of the universe as
a whole. Secondly, inertial  fields are an unremovable reality for an
observer inside a closed universe. Having no possibility to eliminate the
inertial fields by a global transformation of coordinates and time, the
observer is not able in principle to make measurements in such a way that
the chosen RS properties would not affect measurement results. The absolute
ban on such measurements is imposed by the equivalence principle. On the
other hand, in accordance with the operational conception of quantum theory
expressed in its Copenhagen interpretation, a state vector describes
results of measurements carried out on a quantum object under real
conditions created by a certain OM. Therefore, the wave function of the
Universe must carry information on its geometry as well as on a noninertial
RS in which this geometry is studied. It should be expected that QGD of
a closed universe constructed in a mathematically correct manner will
contain the wave functions of this type only. It will be shown in Part II
that QGD of the model Bianchi-IX allowing one to make explicitly
all the calculations confirms the phenomenological predictions based on the
Copenhagen interpretation of quantum theory.

\sect{The operational interpretation\\
of a reference system in a closed universe}
\label{RS_interpret}
According to the above the task of QGD is, firstly, to find an operationally
interpreted gauge-noninvariant wave function of the Universe and, secondly,
to extract information from the wave function about properly geometry of the
Universe as well as in what degree its properties depend on those of OM (a
RS). This approach makes us put a question about a certain physical notion
of an object which one could consider as an OM carrier in a closed universe.

As it was shown by Landau and Lifshitz \cite{Lan}, full information about
dynamical geometry can be obtained directly in experimental way (without
theoretical reconstruction) only in a RS disposed on infinite number of
bodies filling the whole space. Each of the bodies should be equipped by
arbitrarily going clock. {\em The choice of a certain RS} within this class
is realized by {\em choosing defined operations to co-ordinate clocks}
disposed on various bodies. Of course, the equations fixing the choice
(gauge conditions) are noninvariant under the transformations of group of
space-time symmetry (diffeomorphism group) covering all possible reference
systems. It is worth paying attention that a special place of time in the
theory becomes obvious when operationally interpreting the Landau-Lifshitz
RS.

The co-ordination of clocks is performed by setting the metric components
\begin{equation}
\label{g,chi}
g_{0\mu}(x)=\chi_{\mu}\left[\gamma^{ik}(x)\right],
\end{equation}
where $\chi_{\mu}$ is some functional of the 3-metric; $\gamma^{ik}=-g^{ik}$.
A special role of $g_{0\mu}$ in the operational procedure is
bound up with the lack of generalized velocities $\dot{g}_{0\mu}$ in
the gravitational Lagrangian. The latter circumstance indicates that dynamics
of $g_{0\mu}$ is a joint prerogative of a physical object and OM.

It is worth emphasizing that any gauge condition aims at fixing
Eq.\,(\ref{g,chi}). In the classical theory (\ref{g,chi})
can either be given directly before integrating the Einstein equations, or
(when using gauges unsolvable explicitly for $g_{0\mu}$) be found
as a result of integration of the Einstein equations. The choice of a way to
specify a RS does not make real significance since the classical theory is
gauge-invariant by its mathematical structure. The same concerns the quantum
theory of a gauge-invariant S-matrix.

In correspondence with the integrity principle it is necessary to realize the
conception of joint and self-consistent evolution of a physical object and
OM in QGD of a closed universe. The simplest phenomenological hypothesis is
that this evolution is going according to the laws of quantum Hamilton
dynamics. When constructing a gauge-noninvariant QGD this hypothesis becomes
a postulate limiting the class of admissible gauges. A gauge is thought to be
admissible if it extends the phase space of gravitational variables, maintains
the theory being local and does not introduce derivatives of higher orders to
the theory. A gauge condition that satisfies these requirements looks like
\begin{equation}
\label{f,g,chi}
f^{\mu\nu}\dot{g}_{0\nu}=\chi^{\mu},
\end{equation}
where
$f^{\mu\nu}\left(g^{\lambda\sigma}\right),\quad
\chi^{\mu}\left(g_{0\nu},\,g_{0\nu,i},\,\gamma^{ik},\,
\dot{\gamma}^{ik},\,\gamma^{ik},_l\right)$
are algebraic forms of the variables indicated. We shall confine attention
to gauges of the class (\ref{f,g,chi}) because we do not see any technical
possibility to go over to Eq. (\ref{g,chi}) within the framework of the path
integral approach to QGD of a closed universe when gauges unsolvable
explicitly for $g_{0\mu}$, in particular, canonical gauge, are under
consideration.

The gauges (\ref{f,g,chi}) introducing missing generalized velocities to a
Lagrangian, enable us to go over to a Hamiltonian theory in EPS. In such a
theory the indeterminacy principle is valid for all metric components, that,
in turn, allows to deduce a Schr\"odinger equation for the wave function of
the Universe directly from the Hamilton operator equations and commutation
relations. Exactly the same equation can be obtained for the wave function
of the Universe defined through the path integral with the effective action
(\ref{FP_action}) in the class of gauges (\ref{f,g,chi}). For the Bianchi IX
model QGD the equivalence of canonical and path integral approaches in EPS
will be demonstrated in Part II by direct calculations.

Let us turn to discussing a physical nature of an object that is supposed to
be an OM carrier in a closed universe. The conception formulated by Landau
and Lifshitz show that a medium with the following properties must be
considered as a RS in the theory of gravity:
\begin{enumerate}
\item  The medium fills the whole space, i. e. it is continual.
\item A periodic process occurs inside the medium, its characteristic can be
used for choosing metric measurement standards.
\item The symmetry of the medium under diffeomorphism group
transformations is broken.
\end{enumerate}

There is no medium with the enumerated properties in classical physics.
However, one can notice that quantum field theory and particle physics give
examples of objects with similar properties. We mean nonperturbative vacuum
condensate like Higgs or quark-gluon condensates. The vacuum condensate is a
semiclassical medium (though it can have an internal quantum structure); in
the spectrum of its excitations, as a rule, there are (pseudo) Goldstone modes
which, in principle, can be considered as metric measurement standards; at
last, the condensate formation is accompanied by symmetry breaking.

Adoption of the QGD operational interpretation conception allows to predict
that its formalism contains the effect of the origin of a special
gauge-noninvariant gravitational vacuum condensate breaking space-time
symmetry under diffeomorphism group transformations. Various states of the
condensate distinguished by residual symmetry groups correspond to various
Landau -- Lifshitz RS (means of observation upon the Universe as a whole). The
information about physical processes going inside the Universe affects the
quantum-wave Goldstone structures of the excited condensate, i.e. these
structures can be used to record results of measurements.

Appealing to the Landau-Lifshitz RS makes the statement about the
gauge noninvariance of quantum theory of gravity be almost obvious. Indeed,
a formal transformation of coordinates meaning a transition to another gauge,
physically corresponds to removing OM from {\em the whole space} of the
Universe and replacing them by other OM. From the point of view of the
integrity principle that declares existence of unremovable connections between
the properties of an object and OM, it seems to be incredible that such an
operation performed on the {\em whole Universe} scale, would not result in
changing its quantum properties%
\footnote{In the theory of quantum transitions
between asymptotic states it is obvious that the replacement of bodies on
which an asymptotically inertial RS is realized and replacement of detectors
disposed on these bodies cannot change a physical situation. Therefore such a
theory {\em has to be} gauge-invariant.}.

\sect{The wave function of a closed universe and quantization scheme}
\label{scheme}
As was argued in Sec.\,\ref{Asymp.states}, we have no ground to declare
that a wave function of a closed universe must be gauge-invariant.
It was demonstrated in Sec.\,\ref{Integrity},\ref{RS_interpret} that
introducing a wave function containing information about geometry of the
Universe as well as about a RS does not contradict to the Copenhagen
interpretation of QT. Moreover, in Sec.\,\ref{Many-worlds} it was shown that
the theory based on the Wheeler -- DeWitt equation is just a paradigm
which cannot be founded on general QT principles.

With all the above in mind, we cannot require for a wave function of a
closed universe to satisfy the Wheeler -- DeWitt equation. At the same time,
independently on our notion about gauge invariance or noninvariance of the
theory, the wave function has to obey some Schr\"odinger equation. Only
after constructing the wave function of a closed universe satisfying
the Schr\"odinger equation, we shall be able to investigate the question,
under what conditions this wave function could obey the Wheeler -- DeWitt
equation as well.

The Feynman approach contains the procedure of derivation of a Schr\"odinger
equation from a path integral in the Lagrangian form. According to the
physical situation, we consider the path integral without asymptotic
boundary conditions.

In this situation we face the alternative: to work with the path integral
with the Faddeev -- Popov effective action (\ref{FP_action}), or consider
the BFV form of a path integral with following integrating out all momenta
and passing on to a path integral over extended configurational space.
These path integrals should be skeletonized on different gauge-noninvariant
sets of equations that will lead to nonequivalent results. In the present
paper we choose the Lagrangian formulation of the theory based on the action
(\ref{FP_action}) as a starting point for our investigation. Indeed,
it has been already mentioned in Sec.\,\ref{Asymp.states} that for a system
without asymptotic states one cannot declare the BFV effective action
(\ref{BFV_action}) to be BRST-invariant. But the BFV scheme is broken if we
cannot ensure the BRST-invariance of the action: the Fradkin -- Vilkovisky
theorem is not valid then. So there are no grounds to think that the
Hamiltonian formalism for the system without asymptotic states ought to
be constructed along the BFV line. The BFV approach was developed originally
for constructing a relativistic S-matrix of a constrained system. The
purpose of its authors was to build the formalism equivalent to the Dirac
quantization scheme \cite{BFV1}. One should bear in mind that even at the
classical level Dirac's formulation for gravity is a theory the group of
transformations of which does not coincide with gauge group in the
Lagrangian formalism.

Though the path integral approach does not require to construct a
Hamiltonian formulation before deriving the Schr\"odinger equation,
it implies that the Hamiltonian formulation can be constructed. As was
pointed out in Sec.\,\ref{RS_interpret}, it is possible to do it in the
class of gauges (\ref{f,g,chi}). The Hamiltonian can be obtained in a usual
way (according to the rule $H=p\dot q-L$, where $(p, q)$ are the canonical
pairs of EPS) by introducing momenta conjugate to all degrees of freedom
including gauge ones. Thus we get the Hamiltonian formulation in EPS
equivalent to the Lagrangian formulation. We exploit the idea of extended
phase space in the sense that gauge and ghost degrees of freedom are treated
on the equal base with other variables and constrained equations and gauge
conditions will be included to the set of Hamiltonian equations in EPS.
In the second part of our paper we shall consider a model form of the gauge
condition (\ref{f,g,chi}) for the Bianchi IX cosmology. Before quantizing
the model we shall discuss the Lagrangian and Hamiltonian formulations and
compare the latter with the BFV one (see Sec. 10, 11). It will be shown
that Dirac's constraints correspond to a particular form of
gauge-noninvariant Hamiltonian equations in EPS.

The procedure of derivation of the Schr\"odinger equation enables one to
control the correctness of mathematical operations. It is remarkable
that all mathematical expressions appear to be well-definite, in contrast
to attempts of deriving a gauge-invariant Schr\"odinger equation (see
Sec. 14). Such attempts inevitably lead to divergent path integrals, the
divergence being entirely due to the fact that dynamics of gauge variables
is not fixed.

The procedure gives an explicit form of the Hamiltonian operator which
corresponds to the Hamiltonian in EPS (up to the ordering problem). Thus
the Lagrangian and Hamiltonian formulations and the quantization procedure
are consistent in this approach.

As one can see, the ``extended'' set of Lagrangian equations obtained by
varying the effective action (\ref{FP_action}) is of importance in our
consideration. So, before proceeding to our program for the Bianchi IX
model, we shall discuss its properties more profoundly. It will be shown
that the proposed modification of QGD gives rise to the appearance of time
in quantum geometrodynamics.

\sect{The extended set of the Einstein equations\\
and the appearance of time in quantum geometrodynamics}
\label{Lagr.eqs}
Let us consider the gauge
\begin{equation}
\label{gauge}
\partial_0\left(\sqrt{-g}\;g^{0\mu}\right)=0,
\end{equation}
which belongs to the class (\ref{f,g,chi}) and expands the phase space of
gravitational variables. The transition amplitude depends on the action with
gauge-fixing and ghost terms
\begin{equation}
\label{action}
S=\int\Bigl(-\frac1{2\kappa}\sqrt{-g}\,R+\sqrt{-g}\,L_{mat}
  +\lambda_{\mu}\partial_0\left(\sqrt{-g}\ g^{0\mu}\right)
  +\bar{\theta}_{\nu}\hat{M}^{\nu}_{\mu}\theta^{\mu}\Bigr)\,d^4x,
\end{equation}
where $L_{mat}$ is the Lagrangian of nongravitational physical fields;
$$
\hat{M}_{\mu}^{\nu}=
 -\partial_{\mu}\left(\sqrt{-g}\ g^{0\nu}\partial_0\right)
 +\delta_{\mu}^{\nu}\partial_0
  \left(\sqrt{-g}\ g^{0\sigma}\partial_{\sigma}\right)
 +\delta_{\mu}^0\partial_0
  \left(\sqrt{-g}\ g^{\nu\sigma}\partial_{\sigma}\right)
$$
is the Faddeev -- Popov operator (\ref{FP_oper}), corresponding to the gauge
(\ref{gauge}); $\lambda_{\mu}$ are the Lagrange multipliers which will be
further referred to as condensate variables. The variations of the action
(\ref{action}) yields the gauge condition (\ref{gauge}), the ghost equations
\begin{equation}
\label{M,theta}
\hat{M}_{\mu}^{\nu}\theta^{\mu}=0,\quad
\hat{M}^{+\nu}_{\hphantom{+}\mu}\bar{\theta}_{\nu}=0,
\end{equation}
and the gauged Einstein equations
\begin{equation}
\label{R,T}
\frac 1{\kappa}\left(R_{\mu}^{\nu}-\frac 12\delta_{\mu}^{\nu}R\right)=
 T_{\mu(mat)}^{\nu}+T_{\mu(obs)}^{\nu}+T_{\mu(ghost)}^{\nu},
\end{equation}
where
\begin{equation}
\label{Tobs}
T_{\mu(obs)}^{\nu}=-\left(g^{0\nu}\delta_{\mu}^{\sigma}
 +g^{\nu\sigma}\delta_{\mu}^0
 -g^{0\sigma}\delta_{\mu}^{\nu}\right)\partial_0\lambda_{\sigma};
\end{equation}
$$
T_{\mu(ghost)}^{\nu}=
 \bar{\theta}_{\mu,\sigma}g^{\nu 0}\theta^{\sigma}_{,0}
 +\bar{\theta}_{\rho,\sigma}g^{\rho\nu}\delta^0_{\mu}\theta^{\sigma}_{,0}
 -\bar{\theta}_{\sigma,0}\delta^0_{\mu}g^{\rho\nu}\theta^{\sigma}_{,\rho}
 -\bar{\theta}_{\sigma,0}g^{0\nu}\theta^{\sigma}_{,\mu}-
$$
\begin{equation}
\label{Tghost}
 -\bar{\theta}_{\mu,0}g^{\rho\nu}\theta^0_{,\rho}
 -\bar{\theta}_{\rho,0}g^{\rho\nu}\theta^0_{,\mu}
 -\delta^{\nu}_{\mu}\left(
  \bar{\theta}_{\rho,\sigma}g^{\rho 0}\theta^{\sigma}_{,0}
  -\bar{\theta}_{\sigma,0}g^{\rho 0}\theta^{\sigma}_{,\rho}
  -\bar{\theta}_{\rho,0}g^{\rho\sigma}\theta^0_{,\sigma}\right)
\end{equation}
are the quasi-energy-momentum tensor (quasi-EMT) of the OM and ghosts,
respectively. It is supposed, that equations for nongravitational fields
are considered together with (\ref{gauge}), (\ref{M,theta}), (\ref{R,T}).
For the right hand side of (\ref{R,T}) it results from the Bianchi
identities that
\begin{equation}
\label{conservT}
\partial_{\nu}\left(\sqrt{-g}\,T_{\mu}^{\nu}\right)
 -\frac12\sqrt{-g}\;g_{\nu\sigma,\mu}g^{\sigma\lambda}T_{\lambda}^{\nu}=0.
\end{equation}
For $T^{\nu}_{\mu(mat)},T^{\nu}_{\mu(ghost)}$ Eq.\,(\ref{conservT}) holds
identically on the equations of motion; the substitution of the quasi-EMT of
the OM in (\ref{conservT}) enables us to obtain the equations which
demonstrate explicitly the dynamics of the condensate variables:
$$
\partial_{\nu}\left(\sqrt{-g}\,T_{\mu(obs)}^{\nu}\right)
 -\frac{g^{0\lambda}}{g^{00}}\partial_0\lambda_{\lambda}
  \partial_{\mu}\left(\sqrt{-g}\;g^{00}\right)
 =-\partial_{\nu}\left(\sqrt{-g}\;g^{0\nu}\partial_0\lambda_{\mu}\right)-
$$
\begin{equation}
\label{conservTobs}
 -\delta^0_{\mu}\partial_{\nu}\left(\sqrt{-g}\;g^{\nu\lambda}
  \partial_0\lambda_{\lambda}\right)
 +\partial_{\mu}\left(\sqrt{-g}\;g^{0\lambda}\partial_0\lambda_{\lambda}\right)
 -\frac{g^{0\lambda}}{g^{00}}\partial_0\lambda_{\lambda}
  \partial_{\mu}\left(\sqrt{-g}\;g^{00}\right)=0.
\end{equation}

In quantum theory Eqs.\,(\ref{gauge}), (\ref{M,theta}), (\ref{R,T}) are
used for approximating a path integral or considered as operator equations
in the canonical approach. The set of equations has a particular solution,
where
\begin{equation}
\label{theta,lambda}
\theta^{\mu}=0;\quad
\bar{\theta}_{\nu}=0;\quad
\lambda_{\mu}=0;
\end{equation}
\begin{equation}
\label{ReqTmat}
\frac1{\kappa}\left(R_{\mu}^{\nu}-\frac12\delta_{\mu}^{\nu}R\right)
 =T_{\mu(mat)}^{\nu}.
\end{equation}
It is generally accepted that the approximation of a path integral using
(\ref{theta,lambda}), (\ref{ReqTmat}) leads to a gauge-invariant version of
the quantum theory. Correspondingly, in the canonical approach the equations
of motion are the ${k\choose i}$-Einstein equations(\ref{ReqTmat}); all the
other equations make sense only in operating on a state vector, i. e. they
are selection rules for physical states. Actually, as was already mentioned,
one can extract gauge-invariant effects in the theory based on the
gauge-noninvariant set of equations (\ref{gauge}), (\ref{M,theta}),
(\ref{R,T}) as well; to do it when constructing S-matrix it is sufficient
to identify asymptotic states with the vacuum states of ghost, 3-scalar and
3-vector graviton fields. (This procedure is realized in the standard
operator formulation of perturbation theory). Using the particular solutions
(\ref{theta,lambda}), (\ref{ReqTmat}) inevitably makes one perform
mathematically incorrect operations with divergent path integrals.
Thus there remains no ground for using (\ref{theta,lambda}), (\ref{ReqTmat})
except appealing to the strong formulation of the correspondence principle
that is typical for the Dirac approach.

If one puts $\mu = 0$ in the equation (\ref{conservTobs}) for
$T^{\nu}_{\mu(obs)}$ it gives the continuity equation
$$
\partial_{\nu}\left(\sqrt{-g}\,T_{0(obs)}^{\nu}\right)=0,
$$
indicating that when extending phase space a new integral of motion appears
in the physical system,
\begin{equation}
\label{T00eqE}
\int\sqrt{-g}\;T_{0(obs)}^0\,d^3x=
 -\int\sqrt{-g}\;g^{00}\,\partial_0\lambda_0\,d^3x=-E.
\end{equation}
In the gauge (\ref{gauge}) $\sqrt{-g}\;g^{00}$ is a function of space
coordinates only, therefore from (\ref{T00eqE}) one concludes that the
quantity $\partial_0\lambda_0$ contains a constant component
$$
(\partial_0\lambda_0)_{(cond)}=\frac EV,\quad
V=\int\sqrt{-g}\;g^{00}\,d^3x={\rm const},
$$
that we propose to refer to as the energy density of the gravitational
vacuum condensate (GVC). Thus, the existence of the gauge-noninvariant
integral of motion (\ref{T00eqE}) means that in the present theory
gravitational vacuum breaking down the symmetry of the system under
diffeomorphism group transformations shows itself as a real physical
subsystem. Further, the Hamiltonian constraint of the Einstein theory
$H_{(ph)}=0$ is replaced by the constraint $H=E$, $H_{(ph)}$ being a
Hamiltonian of gravitational and material fields and $H$ being a Hamiltonian
in EPS. The constraint $H=E$ after the quantization procedure becomes the
condition on a state vector
\begin{equation}
\label{Hpsi}
H|\psi\,\rangle =-\int d^3x\,\sqrt{-g}\;T_{0(obs)}^0\,|\psi\,\rangle
\end{equation}
instead of the Wheeler -- DeWitt equation (\ref{B-IX_WDW}). The equation
(\ref{Hpsi}) states the spectra coincidence of the Hamiltonian operator in
EPS and the operator
\begin{equation}
\label{intT00obs}
-\int d^3x\,\sqrt{-g}\;T_{0(obs)}^0
\end{equation}
This equation is compatible with the Schr\"odinger equation in EPS and does
not limit the Hamiltonian spectrum by the unique zero eigenvalue. The
comparison with the Schr\"odinger equation allows us to identify the
operator (\ref{intT00obs}) in the Schr\"odinger representation with the
operator $i\partial/\partial t$ (in the Heisenberg representation, as we
saw, (\ref{intT00obs}) is the operator corresponding to the integral of
motion). We shall refer to the tensor $T^{\nu}_{\mu(obs)}$ as a quasi-EMT
of the condensate.

The problem of time, so typical of the Wheeler -- DeWitt QGD, does not arise
in the proposed modification of QGD; there is no necessity to introduce time
in an unnatural way, by declaring a scale factor or a scalar field to be a
parameter replacing a time variable (see, for example, \cite{Vil,Cast}).
The proposed formulation of QGD, however, enables not just to avoid the
problem of time, but to point out the source of time appearance in QGD. The
obtained result shows that the origin of time as a physical concept in a
quantum closed universe is bound up with an observer's presence in it. The
time appears owing to our pointing an instrument to measure it, introducing
into consideration a subsystem of the Universe -- the GVC, that could be
regarded as the carrier of OM (a reference system). In principle, we get the
possibility to examine the Universe evolution in time considering sequential
quantum transitions between states with certain eigenvalues of the operator
(\ref{intT00obs}), the quantity $E$, defining a full energy of the GVC in a
closed universe. The problem of time in the Wheeler -- DeWitt QGD, in our
opinion, results from using incorrect mathematical procedures, breaking a
quantum integrity of the system ``a physical object + OM".

The equation (\ref{conservTobs}) reveals in the GVC appropriated to the
gauge (\ref{gauge}) the presence of both mentioned above components which are
necessary for performing the function of a RS: a continual component breaking
the symmetry under diffeomorphism group and a periodic one that can give
metric measurement standards. It results from the fact that the equation
(\ref{conservTobs}) for the parameter $\partial_0\lambda_{\sigma}$ of the
quasi-EMT $T^{\nu}_{\mu(obs)}$ after substituting
$\partial_0\lambda_{\sigma}=\partial_{\sigma}\chi$
can be reduced to the d'Alembert equation
\begin{equation}
\label{chi_eqn}
\partial_{\mu}\left(\sqrt{-g}\;g^{\mu\nu}\partial_{\nu}\chi\right)=0,
\end{equation}
its general solution containing both components, a constant and wave one 
interacting with the metric.
	
Up to some extent of accuracy, the quantum field $\chi$ can be interpreted as
a Goldstone mode of the GVC collective excitations. It is worth noting that 
the association of the field $\chi$ precisely with solutions to the d'Alembert 
equation (\ref{chi_eqn}) describing massless particles propagating with the 
speed of light is a direct consequence of our choice of the concrete gauge 
(\ref{gauge}). At all the other gauges extending phase space objects like the 
GVC and its collective excitations would appear as well. However, the latter 
ones would not propagate with the speed of light; in some cases their 
velocities are less than the speed of light, and superlight in other cases. 
We do not know the principles of integrated system formation; probably,
the properties of the field $\chi$ reflected in Eq.\,(\ref{chi_eqn}) might 
serve as a phenomenological criterion in this situation. It is easy to see 
from Eq.\,(\ref{conservTobs}) that the interactions between the field $\chi$ 
quanta and the ordinary matter are as weak as similar interactions of 
gravitational waves. Exactly for this reason the wave structures of a field
$\chi$, in principle, may be used for the registration of information about 
the Universe evolution. The phenomenological character of the discussed 
theory and the lack of physical principles regulating formation of the 
integrated system ``a physical object + OM", of course, do not allow us to 
put the question about the concrete mechanism of the information registration. 
This problem, seemingly, is a prerogative of a future theory.


\small
\begin{thebibliography}{99}
\bibitem{DeWitt}  B. S. DeWitt, {\it Phys. Rev.\/} {\bf 160}, 1113 (1967).
\bibitem{BV}      I. A. Batalin and G. A. Vilkovisky, {\it Phys. Lett.\/}
                  {\bf B 102}, 27 (1981); {\it in}: Proceed. of the 2nd
                  Moscow Seminar on Quantum Gravity, W. Scientific,
                  Singapore, 1982.
\bibitem{BFV1}    E. S. Fradkin and G. A. Vilkovisky, {\it Phys. Lett.\/}
                  {\bf B 55}, 224 (1975).
\bibitem{BFV2}    I. A. Batalin and G. A. Vilkovisky, {\it Phys. Lett.\/}
                  {\bf B 69}, 309 (1977).
\bibitem{BFV3}    E. S. Fradkin and G. A. Vilkovisky, CERN Report TH-2332,
                  1977.
\bibitem{BFV4}    E. S. Fradkin and T. E. Fradkina, {\it Phys. Lett.\/}
                  {\bf B 72}, 343 (1978).
\bibitem{Duncan}  M. J. Duncan, {\it Preprint} UMH-TII-916/90.
\bibitem{Everett} H. Everett, {\it Rev. Mod. Phys.\/} {\bf 29}, 454 (1957).
\bibitem{Wheeler} J. A. Wheeler, {\it ibid.}, 463.
\bibitem{BP}      A. O. Barvinsky and V. N. Ponomariov,
                  {\it Izvestiya Vuzov, Fizika\/} {\bf 3}, 37 (1986).
\bibitem{Lan}     L. D. Landau and E. M. Lifshitz, ``Theory of Fields",
                  Nauka, Moscow, 1988.
\bibitem{HP}      S. W. Hawking and D. N. Page, {\it Nucl. Phys.\/}
                  {\bf B 264}, 185 (1986).
\bibitem{Di}      P. A. M. Dirac, {\it Proc. Roy. Soc.\/} {\bf A 246}, 333
                  (1958).
\bibitem{ADM}     R. Arnowitt, S. Deser and C. W. Misner, {\it Phys. Rev.\/}
                  {\bf 117}, 1595 (1960).
\bibitem{Fad}     L. D. Faddeev, ``The Hamiltonisn Form of the Theory of
                  Gravity", {\it in}: Theses of the 5th International
                  Conference on Gravitation and Cosmology, Tbilisi, 1968.
\bibitem{FP1}     L. D. Faddeev and V. N. Popov, {\it. Phys. Lett.\/}
                  {\bf B 25}, 30 (1967).
\bibitem{FP2}     L. D. Faddeev and V. N. Popov, {\it Uspekhi Phys. Nauk\/},
                  {\bf 111}, 427 (1973).
\bibitem{Hennaux} M. Hennaux, {\it Phys. Rep.\/} {\bf 126}, 1 (1985).
\bibitem{Bohr1}   N. Bohr, ``Atomic Physics and Human Knowledge", N.Y., 1958.
\bibitem{Bohr2}   N. Bohr, ``Quantum Physics and Philosophy", {\it in}:
                  Philosophy in the Mid-Century. A Survey, Firenze, 1958.
\bibitem{Vil}     A. Vilenkin, {\it Phys. Rev.\/} {\bf D 33}, 3560 (1986).
\bibitem{Cast}    M. Castagnino, {\it Phys. Rev.\/} {\bf D 39}, 2216 (1989).

\end{thebibliography}
\end{document}